\documentclass[12pt]{iopart}
\usepackage{graphicx, epsfig}
\usepackage{SIunits}

\begin{document}
\title{Capped colloids as light-mills in optical traps}
\author{F S Merkt, A Erbe and P Leiderer}
\address{Universit\"{a}t Konstanz, Universit\"{a}tsstr. 10, 78457 Konstanz}
\ead{florian.merkt@uni-konstanz.de}
\date{\today}

\begin{abstract}
Custom-designed colloidal particles in an optical tweezers act as
light-mills in a fluid. In particular, aqueous suspensions of capped
colloids, in which half of the surface is covered with metal layers,
are investigated. Due to their asymmetry, the capped colloids can
act as rotators when exposed to intense laser fields. Particles of
\unit{4.7}{\micro\meter} in diameter are observed rotating around
the focus of a laser beam. For low intensities, particles become
trapped close to the spot of highest laser intensity. Above a
threshold value of about \unit{4}{mW} in total beam intensity, the
particles move away from the center of the focus and start to rotate
at frequencies of about \unit{1}{Hz}. The balance of forces due to
light pressure and hydrodynamic forces gives a constant rotation
rate. The speed of the spinning particle increases linearly with
laser power to above \unit{2}{Hz} until the particles are ejected
from the focus for intensities higher than \unit{7}{mW}. Magnetic
caps introduce further possibilities to tune the rotation rates.
\end{abstract}

 \submitto{\NJP} \maketitle

\baselineskip12pt
\section{Introduction}
Laser tweezers allow microscale particles to be positioned with high
precision and reproducibility. The main advantage of this technique
is its versatility. It can be performed in almost any transparent
medium, especially in cells containing a liquid. Thus, a large
number of systems~(ranging from biological samples such as cells to
colloidal particles) can be manipulated using this technique. A
well-defined rotation of the particles can be used to further
investigate particles on the micron and even nanometer scale.
Therefore a large number of experiments has been performed during
the past years in order to investigate the rotational motion of such
particles. Typical approaches used the absorption of polarized light
that transfers spin or angular momentum and refraction of the beam
by custom-made objects. Rotation around the beam axis
\cite{Galajda:2001,Higurashi:1994} and perpendicular to it
\cite{Higurashi:1998} could be studied and theoretically explained
\cite{Gauthier:1995}. Additionally, turning the angle of
polarization of an optical tweezers can also lead to rotation of
birefringent particles \cite{Friese:1998}. Wherever light pressure
propels the motion of particles, the rotating object needs to be
asymmetric. In the system investigated here, asymmetry is introduced
by coating one side of the spheres with metal caps. Visualization of
Brownian motion of similar particles \cite{Behrend:2004a} and
rotation of particles with magnetic caps in a magnetic field have
already been demonstrated \cite{Behrend:2004b}. Accordingly, charged
particles with metal caps move in an electric field
\cite{Takei:1997}. More detailed studies on rotation in optical
traps use biochemically joined particles as dimers \cite{Luo:2000}
and pieces of glass powder \cite{Yamamoto:1995} as rotating agents.
In this work, we describe the controlled off-axis rotation of capped
colloids around a Gaussian laser spot is described.\\

\section{Sample preparation and experimental setup}
The samples, named capped colloids, were prepared in the process
illustrated in Figure~1. It starts with the production of a
monolayer, which is similar to the method described in
\cite{Burmeister:1998}. A drop of suspension with
\unit{4.7}{\micro\meter} Silica particles (SS05N from Bangs
Laboratories) totally wets a thoroughly cleaned borosilicate glass
cover slip. It dries in a cell at water saturation pressure, which
is tilted by about \unit{5}{\degree}. Through self-assembly during
evaporation of the liquid, the particles arrange in hexagonal order.\\
The sample is then transferred into an evaporation chamber and a
coating is added. After an adhesive layer of \unit{1-2}{\nano\meter}
Ti, various metals such as Ni, Au or Co-Pd multilayers are deposited
on the particles \cite{Albrecht:2005}. The final coating, added by
e-beam evaporation, consists of a \textrm{SiO$_{\textrm{x}}$} layer,
which should guarantee uniform surface properties such as surface
charge.\\
To detach the particles from the substrate, a stepper motor stage
dips the cover slips into a water basin under \unit{45}{\degree} at
about \unit{10}{\micro\meter\per\second}. Surface tension peels the
colloidal particles off the substrates so that they float on top of
the water surface \cite{Burmeister:1999}. After pipetting the excess
fluid from the reservoir, the particles are mixed with the residual
liquid. Using this recipe, they disconnect from each other and give
suspensions of variable concentration in capped colloids.
Alternatively, ultrasound also removes particles from the surface
\cite{Agayan:2004}.\\
Finally, the suspensions are confined between two glass plates
separated by an O-ring. The lower glass slip is coated with a PMMA
layer to provide a smooth surface, which prevents sticking of the
particles. For the same reason, the particles are suspended in
deionized water.\\
Such sample cells are examined in an optical tweezers setup. The
colloidal spheres are imaged with an inverted video microscope onto
a CCD camera, and particle positions are determined with a rate of
\unit{1}{Hz} and a lateral accuracy of \unit{100}{\nano\meter}. From
these data we obtain particle positions and the relative
orientations of the caps.\\
A frequency doubled Nd:YVO4 laser ($\lambda =
\unit{532}{\nano\meter}$) generates the Gaussian shaped laser focus
with a full width at half maximum of about \unit{2.8}{\micro\meter}.
The light is collimated through a 20x objective placed directly
above the sample cells. Rotation rates were determined with a photo
diode mounted at one side of the sample cell by recording the
sequence of light pulses.\\

\section{Results}
The orientation and rotational motion of capped colloids are clearly
visible. This allows to characterize the system in a controlled and
precise way.\\
Figure~2 depicts capped colloids rotating in the optical trap. The
laser beam driving the motion of the particles is reflected by the
inner and outer cap surfaces. The light-house effect results from
this reflection at the cap. It becomes visible in the images because
of the scattering of the reflected light from the substrate surface.
Figure~2a and a video in the supporting material visualize this
effect. In order to determine the orientation of the cap in greater
detail, the intensity of the laser light in front of the camera is
reduced. This reveals that the motion of the particle is eccentric.
It moves on a circle with a diameter of about
\unit{1.5}{\micro\meter} around the focal point. Although the exact
orientation of the cap is not immediately obvious, it is inferred
from the images that its edge aligns vertically. So the plane
separating the coated and uncoated parts lies perpendicular to the
substrate. Tilting a cap away from this position lowers the rotation
rate until the particle stops. This can be achieved by an external
field interacting with a magnetic cap. There is no preference either
for rotation in clockwise or counterclockwise directions and jumps
between them occur spontaneously.\\
To induce the rotation of the particles, a threshold value in laser
intensity needs to be overcome. Below this intensity, the capped
particles can be trapped and moved, but do not rotate. As shown in
Figure~3, above this value the rotation rate increases almost
linearly with the laser power. Each individual particle shows a
monotonous increase in rotation rate with laser power. The errorbars
indicate the distribution of the frequencies for an ensemble of
particles. The standard deviation in rotation rate for one specific
capped colloid is approximately the symbol size. At a certain point
(here $\sim~7.5~mW$), the light pressure is high enough to expel
particles from the laser focus.\\
In further investigations, the mechanism of the rotation was
investigated in more detail. In particular, the relative positions
of the laser tweezers and the particle center have been determined.
At first, the position of the laser focus was determined from a
picture without the particle. Subsequently, colloids and caps were
located in images resulting after a threshold in brightness level
had been applied. Dark areas corresponding to particles in the
pictures were identified and located by edge detection. The position
of the cap was found as the barycenter of the dark region. With this
information, the normal vector of the caps could also be determined.
It was given by the relative position of the particle center and the
barycenter of the cap. For each particle the distance to the beam
center was determined. Finally, the relative coordinates were
converted into the coordinate system of the capped colloid. The
resulting distribution of focal positions is depicted in
Figures~4c,~f.\\
An effective potential for the asymmetric particles can be given and
illustrates the bistability of the system. The Boltzmann equation is
used to convert the frequency of relative positions as displayed in
Figures~4a, d into a potential energy landscape. Only the radial
dependence of the effective potential is retained. The angle between
the cap and the vector from the particle center to the position of
the laser focus has been integrated out. For uncapped particles, the
potential has a minimum at the origin and increases with radius as
shown in \cite{Mangold:2003}. For capped particles, the pivotal
minimum in the radially averaged potential shifts to values around
\unit{0.5}{\micro\meter} as seen in Figure~4b. Here, particles are
simply held in the tweezers. The potential further develops a second
minimum for higher laser powers as in Figure~4e. For increasing
intensity, the dip at a distance of \unit{1.5}{\micro\meter} from
the center of the laser focus becomes more and more pronounced. When
the particles are trapped in this state, they rotate around the focus.\\
The orientation vectors for the caps always point against the
direction of the motion. Brownian motion can lead to spontaneous
changes in direction from clockwise to counterclockwise or vice
versa. In the reference frame of the particles in Figure~4f, this
corresponds to a position of the focus which is more to the left or
more to the right of the center, respectively. The points close to
the middle correspond to usual tweezing as in the case of lower
intensity in Figure~4c. The ring-like distribution of the dots is
due to the limited resolution of the camera, which leads to an
artificial quantization of the relative positions of the particles.\\
Another approach to find out more about the behaviour of capped
colloids in optical traps is to use magnetic caps. All capped
colloids rotate irrespective of the intrinsic magnetization in the
caps. Without a magnetic field, we found no difference in the
interaction of the laser with non-magnetic colloids. In the absence
of a magnetic field, particles having an in-plane magnetization from
a \unit{50}{\nano\meter}~Ni cap or a perpendicular magnetization
from CoPd-multilayers rotate as expected. When a magnetic cap is,
however, tilted in the beam direction by an external field from two
Helmholtz coils, the rotation slows with increasing magnetic field
strength. Small magnetically stabilized clusters or short chains of
particles can be rotated as well. Often CoPd-covered particles form
ensembles of three particles so that their caps are pointing toward
their center. Such clusters with a larger hydrodynamic radius rotate
more slowly than individual particles. The reflectivity of the caps
can be tuned by layer depth and choice of material. A minimum layer
depth of about \unit{20}{\nano\meter} of metal coating is needed to
obtain a significantly reduced transmissivity of the capped side.
Particles with thinner caps do not rotate.\\

\section{Discussion and comparison with similar systems}
From a comparison with experiments on specially designed particle
dimers in \cite{Luo:2000}, it is assumed that light pressure drives
the system. Employing capped colloids as rotators, however,
simplifies the geometry of the scattering object. For capped
colloids, several forces contribute to the proposed rotation
mechanism. Due to the scattering force, their transparent part is
drawn into the focus of the laser tweezers. On the other hand, the
momentum transfer from internal and external reflections at the
coated surface pushes the cap. Both components balance at an
equilibrium distance from the beam center. This leads
to the second minimum in the effective potentials.\\
Furthermore, a torque on the particle results because of the
asymmetric orientation of the cap towards the beam axis. Therefore,
the angular velocity increases until it is balanced by hydrodynamic
forces. Due to this asymmetric position, higher torques can be
exerted compared to particles rotating about an axis through their
center. To estimate the net torque from the viscous drag on the
particle, simplifications in analogy to \cite{Luo:2000} yield a
value of about \unit{2.4\cdot10^{-18}}{Nm}.\\
The rotation of capped colloids is robust for various choices of
inner cap material (Al, Au and thin layers of C), given that Au and
{\rm SiO} always cover the outer surface. This supports the
intuitive conclusion in \cite{Luo:2000} that the light pressure due
to reflection at the outside shell drives the motion.\\
During rotation, irregularities in the particle motion occur. Since
Brownian motion leads to a change in the direction of rotation of
the particles, it might also generate fluctuations in the rotation
rates. For constant laser intensity, it is expected that smaller
particles will rotate faster at fixed laser intensity, as is
reported for chunks of glass powder \cite{Yamamoto:1995}. Contrary
to capped colloids, however, the size and shape of such irregular
particles is hard to characterize.\\
The manipulation of rotating capped colloids with a magnetic field
shows that indeed the scattering geometry determines the rotation
rate. It can be used to adjust the orientation of the particles
during rotation. The more the cap is tilted, the slower does the
particle rotate.

\section{Conclusions and Outlook}
In summary, capped colloids rotate around the focus of an optical
tweezers with tunable angular velocity. Above a threshold, the
particles rotate eccentrically with a rotation rate proportional to
the laser intensity. Rotators can be fabricated by floating the
capped particles off the substrates after evaporation of the metal
layers. This prevents damage due to ultrasound and may facilitate
treatments such as biofunctionalization. The rotators can be moved
around and actuated by a simple optical trap. Light pressure pushes
the particle in an equilibrium distance from the beam center.
Effective potentials are derived from the relative positions of the
trap and the particle. They visualize the transition from normal
tweezing to rotation.\\
This design allows to exert torques for pumps, motors or drills on
the micron-scale. Making use of the magnetic properties of the
particles may become a further step towards micro-machines. A
feedback loop for the laser power could be used to further control
the rotation speed as in \cite{Yamamoto:1995}. Additionally, a
magnetic field can set the orientation of magnetic caps.

\ack We thank K. Mangold, M. K\"{o}ppl and L. Baraban for helpful
discussions and support in developing the sample preparation and
measurements, L. Kukk and H. Ballot for technical advice and
assistance. We gratefully acknowledge financial support from the
Deutsche Forschungsgemeinschaft (DFG) through SFB 513.

\begin{figure}[ht]
  \centering
  \caption[preparation of aqueous suspensions of capped colloids]{\textmd{\textbf{Preparation of capped colloids.}~~\textbf{a} A \unit{30}{\micro\liter} drop of a \unit{10}{\%} solids stem suspension of \unit{4.7}{\micro\meter} Silica particles dries on a clean cover slip. The particles arrange into a monolayer and diverse metals are evaporated on top. After dipping into a water reservoir the layer of particles detaches from the substrate and floats on the fluid surface. Finally, stirring breaks the particle bonds and the suspension is enclosed in transparent samples cells.~~\textbf{b} SEM image of two \unit{4.7}{\micro\meter}~Si particles on borosilicate glass object slides that have been coated with \unit{50}{\nano\meter}~Ni, \unit{50}{\nano\meter}~Au and \unit{20}{\nano\meter}~\textrm{SiO$_{\textrm{x}}$} layers. }}
  \label{fig:preparation}
\end{figure}

\begin{figure}[ht]
  \centering
  \caption[rotation of capped colloids in an aqueous suspension]{\textmd{\textbf{Counterclockwise and clockwise rotation of capped colloids around a laser focus.}~~\textbf{a} \unit{4.7}{\micro\meter} Silica colloids with a \unit{50}{\nano\meter}~Au and \unit{20}{\nano\meter}~\textrm{SiO$_{\textrm{x}}$} cap rotate counterclockwise around the focus of a laser beam of \unit{5.44}{mW} in an aqueous suspension. The inner cap surface reflects the light of the optical trap similar to a mirror reflecting a signal fire in a light-house. Additionally, a supplementary 1.3 MB mpeg1 is available online.~~\textbf{b} Filtering out most of the laser light at \unit{4.25}{mW} for a clockwise rotating particle reveals that it is held eccentrically. The transparent half without cap points into the direction of motion. In both cases, the time difference between successive pictures is approximately \unit{0.3}{s}.}}
  \label{fig:rotation}
\end{figure}

\begin{figure}[ht]
  \centering
  \caption[rotation rate over trap intensity and over magnetic field]{\textmd{\textbf{Onset and frequency of rotation.}~~Below a threshold of \unit{4}{mW} of total laser intensity in the optical tweezers \unit{50}{\nano\meter}~Au and \unit{20}{\nano\meter}~\textrm{SiO$_{\textrm{x}}$} capped colloids can be trapped. Above this rotation sets in until for intensities higher than \unit{7.5}{mW} the light pressure expels the particles from the laser focus.}}
  \label{fig:rates}
\end{figure}

\begin{figure}[ht]
  \centering
  \caption[visualization of stabilization of magnetic capped colloids]{{\textbf{Relative positions of laser focus and particle center.}~~\textbf{a} For \unit{2.80}{mW} laser intensity the capped colloidal particle stays close to the focus of the optical tweezers.~~\textbf{b} Since the successive positions follow a Boltzmann distribution, a radially averaged effective potential can be derived.~~\textbf{c} Seen in the reference frame of the colloidal particle, the laser spot stays close to the center.~~\textbf{d} At \unit{5.44}{mW} particles are expelled from the center and rotate at a finite distance about the beam axis.~~\textbf{e} The potential shows a global minimum corresponding to rotation, while the small dip identifies normal tweezing. Typical error bars were estimated.~~\textbf{f} The rotation is counterclockwise for a position of the focus on the left side and clockwise for positions right to the center. The limited resolution of the optical microscope causes the grid-like distribution of data points in panels \textbf{a}, \textbf{d}, and the circular artifacts in \textbf{c}.}}
  \label{fig:stabilization}
\end{figure}

\end{document}